\documentclass[aps,prd,twocolumn,nofootinbib,superscriptaddress]{revtex4-1}
\usepackage{amssymb,amsmath,amstext,amsfonts,color,ulem,textcomp,amsthm}
\usepackage{graphicx,multirow,xcolor}

\newcommand{\beq}{\begin{equation}}
\newcommand{\eeq}{\end{equation}}
\newcommand{\bal}{\begin{aligned}}
\newcommand{\eal}{\end{aligned}}
\def\d{\mathrm {d}}

\begin{document}
	
\title{Geodesic completeness of anisotropic cosmologies and the null energy condition} 
	
\author{Sebastian Garcia-Saenz}
\email{sgarciasaenz@sustech.edu.cn}
\affiliation{Department of Physics, Southern University of Science and Technology, Shenzhen 518055, Guangdong, China}

\author{Junjie Hua}
\email{junjiehua@sjtu.edu.cn}
\affiliation{Department of Physics, Southern University of Science and Technology, Shenzhen 518055, Guangdong, China}
\affiliation{Department of Astronomy, School of Physics and Astronomy, Shanghai Jiao Tong University, Shanghai 200240, China}

\author{Abbas M. Sherif}
\email{abbasmsherif25@gmail.com}
\affiliation{Institute of Mathematics, Henan Academy of Sciences (HNAS), 228 Mingli Road, Zhengzhou 450046, Henan, China}

\begin{abstract}
We demonstrate that a Bianchi-I spacetime in general relativity must be geodesically past-incomplete under the assumption of the null energy condition and provided there exists a time in which the universe is expanding in all directions. Spacetimes which are always contracting along at least one direction are not subject to the conclusion, and we provide an explicit example. A distinguishing feature of our theorem is that it makes no assumption on the global topology of the spacetime.
\end{abstract} 

\maketitle

\section{Introduction} \label{sec:intro}
	
The breakdown of general relativity (GR) at short distances or high energies is perhaps our best guide towards a theory of quantum gravity. This is clearly exemplified by the wealth of theoretical developments achieved in the context of black holes (see e.g.\ the reviews~\cite{Harlow:2014yka,Raju:2020smc,Chen:2021lnq}), for which GR predicts the existence of a curvature singularity and hence a breakdown of the classical spacetime description. It is natural to expect that cosmological spacetimes should provide a similarly fruitful arena to study fundamental questions about gravity. To this end, one would like to first fully understand the conditions under which GR indeed fails to provide a complete description. One general and well-defined way to pose this question is to ask whether the spacetime is geodesically complete given some energy conditions on the matter content.

This question has of course a long history, starting with the celebrated `singularity' theorems of Hawking and Penrose~\cite{Hawking:1965mf,Hawking:1966vg,Hawking:1966sx,Hawking:1966jv,Hawking:1967ju,Hawking:1970zqf}.\footnote{We recall that the Hawking-Penrose theorems strictly speaking are not about singularities but rather about geodesic incompleteness.} The main virtue of these results is that, similarly to Penrose's original theorem~\cite{Penrose:1964wq}, they do not rely on symmetry assumptions, thus significantly extending elementary results based on the Friedmann-Lema\^itre-Robertson-Walker (FLRW) metric. They do however require the strong energy condition (SEC) on the energy-momentum tensor, which is violated already by standard matter fields at the classical level. From a physics perspective, it is then clearly pertinent to ask if and when the assumption on the SEC can be relaxed.

In this paper we make progress on this question. Instead of the SEC, we adopt the null energy condition (NEC) as our main assumption, which is indeed more physical knowing that its violation is generically associated with fundamental pathologies~\cite{Hsu:2004vr,Dubovsky:2005xd,Buniy:2005vh,Buniy:2006xf,Sawicki:2012pz,Rubakov:2014jja,Libanov:2016kfc}.\footnote{As is well known, the NEC is actually violated at the quantum level. We will comment on this point in Sec.\ \ref{sec:discussion}.} Moreover, an averaged version of the NEC can in fact be proved in full generality within quantum field theory~\cite{Wald:1991xn,Flanagan:1996gw,Faulkner:2016mzt,Hartman:2016lgu,Balakrishnan:2017bjg}. In the context of the FLRW metric, it is then not hard to demonstrate that the NEC implies that the spacetime must be geodesically past-incomplete if it is expanding~\cite{Garcia-Saenz:2024ogr}. Here we consider instead the Bianchi-I class of metrics as a simple yet physically relevant example of an anisotropic cosmological spacetime. Our main result is that the same incompleteness theorem applies in this case, with an added assumption on the expansion: namely, that there exists a time in which the universe is expanding in all directions. Interestingly, the theorem does not hold without this assumption, and we provide an explicit counter-example.

The question on the completeness of Bianchi-I spacetimes is physically and historically interesting. The famous Kasner solution~\cite{Kasner:1921zz}, which is the unique vacuum Bianchi-I solution of the Einstein field equations, is conjectured to provide a universal description of spacetime in the vicinity of a curvature singularity, forming the building block of the so-called Mixmaster model~\cite{Misner:1969hg,Belinsky:1970ew,Belinsky:1982pk} or as the attractor phase in non-oscillatory models~\cite{Eardley:1971nj,Barrow:1978hgi,Isenberg:1989gq,Berger:2002st}. The relationship between Bianchi-I metrics and spacetime singularities thus appears as particularly important, perhaps even beyond the context of cosmology.

Beyond classical general relativity, Bianchi-I spacetimes have also emerged as a primary testing ground for quantum gravity, particularly in Loop Quantum Cosmology (LQC). A major achievement of LQC has been the resolution of strong curvature singularities in various Bianchi models, including type I~\cite{Gupt:2012aa,Corichi:2009aa}. However, the distinction between strong and weak singularities has proven crucial: while strong singularities are often resolved, weak singularities---where geodesics may still be extendable---can persist in some models~\cite{Saini:2019aa}. Moreover, the outcome can depend sensitively on the quantization scheme and on the imposition of energy conditions~\cite{Corichi:2009aa,Saini:2017bb}. Our classical theorem thus provides an important baseline: it identifies precisely the conditions under which the classical spacetime is geodesically incomplete, thereby clarifying what any viable quantum theory must address.

The question of geodesic completeness in Lorentzian geometry has been extensively studied from a mathematical perspective, with fundamental results relating curvature bounds to completeness (see e.g.~\cite{Beem:1996bk,Sanchez:1997aa,Ecker:2008aa} for reviews). The classical Hopf-Rinow theorem, which guarantees completeness for Riemannian manifolds, has no direct analogue in the Lorentzian setting due to the indefinite metric. Instead, one must rely on energy conditions or curvature constraints to derive incompleteness. Our main result contributes to this broader mathematical program by establishing a sharp condition for Bianchi-I spacetimes, a class that has received attention in the mathematical relativity literature due to its homogeneity~\cite{Tanimoto:2003aa,Randall:2005aa}.

It is worth commenting on previous results concerning the link between the NEC and geodesic completeness of general cosmological spacetimes. First, the well-known Borde-Guth-Vilenkin (BGV) theorem~\cite{Borde:2001nh} (see~\cite{Borde:1993xh,Borde:1994ai,Borde:1996pt} for earlier related work and~\cite{Conroy:2014dja,Vilenkin:2014yva,Kothawala:2018ghr,Kinney:2021imp,Kinney:2023urn,Pavlovic:2023mke,Lesnefsky:2022fen,Easson:2024uxe,Easson:2024fzn,Garcia-Saenz:2024ogr,Kinney:2026kvw} for recent refinements and applications) is in fact even more general as it does not assume any energy condition. The problem however is that the BGV theorem does not always provide an easy-to-use diagnostic tool: one needs to identify a geodesic congruence for which the average expansion is positive. In the case of FLRW one has the comoving geodesics as a natural choice, but such obvious candidates will not necessarily be available in more general spacetimes. Our theorem circumvents this difficulty by using the NEC as a substitute: rather than requiring a congruence with positive average expansion, we assume only that the universe is expanding in all directions at the present time, a condition that is mathematically tractable. As we show in Section IV, this distinction is crucial: the BGV theorem does not imply a violation of the NEC in Bianchi-I, because the three axes need not have overlapping intervals of positive expansion. Our counter-example in Sec.\ \ref{sec:counterexample} explicitly demonstrates this. The difficulties in diagnosing past-incompleteness from BGV and the NEC have been recently discussed in detail in~\cite{Kinney:2026vyg}. Second, there have been concrete attempts to generalize Hawking's theorem~\cite{Galloway:2017mts,Lesourd:2018vrr,Ling:2025slu,Galloway:2025bth}. The available results still do require additional assumptions on the global topology or some global focusing condition, which are not always linked with physics in any obvious way, as is the case with the NEC.\footnote{On a separate line of research going back to the work of Yau~\cite{Yau1982}, there exists a significant number of important results concerned with the `rigidity' of the Hawking-Penrose theorems, i.e.\ the problem of classifying the special cases that avoid the conclusion of the theorem by breaking some of the technical assumptions~\cite{Bartnik1988,Eschenburg1988,Galloway1989,Newman1990,Galloway:2012aa,Galloway:2016uix}.} We therefore think our theorem, although significantly more modest in scope, is of interest due to its direct connection with plausible physical conditions.

\section{Elements of Bianchi-I spacetimes} \label{sec:bianchi}

In this section we derive some elementary facts about Bianchi-I spacetimes in the context of GR. We work in $3+1$ spacetime dimensions, although all the results straightforwardly generalize to arbitrary dimension. The reader may find more in-depth expositions for instance in~\cite{Ellis:1998ct,Pereira:2007yy,Arefeva:2009aa}; see also the Appendices for additional details.

In a suitable gauge, the line element of the Bianchi-I spacetime is given by
\beq
\d s^2=-\d t^2+\sum_{i=1}^3 a_i^2(t)(\d x^i)^2 \,.
\eeq
Throughout this paper we assume the scale factors $a_i$ are smooth positive functions in the domain $(t_{\rm i},t_{\rm f})$ (with $t_{\rm i}=-\infty$ as a necessary condition for past completeness). For each spatial direction, we define a Hubble parameter $H_i(t):= \dot{a}_i/a_i$. The sum of these defines an expansion scalar,
\beq
\theta(t):=\sum_{i=1}^3 H_i(t) \,.
\eeq
One may also define an average scale factor as $a(t):=[a_1a_2a_3]^{1/3}$, with corresponding Hubble parameter $H(t):=\dot{a}/a$. It then follows that $\theta=3H$.

We assume the Einstein field equations hold. The isometries of the spacetime then imply that the energy-momentum tensor has the form $T^{\mu}{}_{\nu}={\rm diag}\left(-\rho,P_1,P_2,P_3\right)$, where we interpret $\rho$ as an energy density and $P_i$ as a pressure component. It may then be proved (see Appendix \ref{app:general NEC} for details) that the NEC is equivalent to the conditions
\beq
\rho+P_i\geq0 \quad \forall\,i \,,
\eeq
or, more simply, $\rho+\min\left(P_1,P_2,P_3\right)\geq0$. Using the Einstein equations these may be rewritten as inequalities involving the $H_i$,
\beq \label{eq:H_i ineqs from NEC}
-\dot{H}_{i+1}-\dot{H}_{i+2}+H_i(H_{i+1}+H_{i+2})-H_{i+1}^2-H_{i+2}^2\geq 0 \,,
\eeq
which should be understood in a cyclic sense, i.e.\ with the convention $H_{4}\equiv H_1$ and $H_{5}\equiv H_2$. One also has, from the trace of the Einstein equation,
\beq
\frac{1}{M_P^2}\left[\rho+\frac{1}{3}\sum_{i=1}^3 P_i\right]=-2\dot{H}-\sigma^2 \,,
\eeq
where $M_P\equiv (8\pi G)^{-1/2}$ is the Planck scale, and
\beq
\sigma^2:=\sum_{i=1}^3 (H_i-H)^2 \,,
\eeq
defines a shear scalar.\footnote{Notice that here $\theta$ and $\sigma^2$ are simply definitions and need not correspond to the expansion and shear of an actual geodesic congruence. They are convenient definitions for the reason that, in the isotropic or FLRW limit, they agree with the expansion and shear scalars of a comoving geodesic congruence.} Given that $\frac{1}{3}\sum_{i=1}^3 P_i \geq \min\left(P_1,P_2,P_3\right)\geq0$, we conclude that the NEC implies
\beq
\dot{H}(t)\leq 0 \,.
\eeq

\section{Null energy condition and past incompleteness} \label{sec:theorem}

Our main result is that the NEC, specifically the inequalities in Eq.\ \eqref{eq:H_i ineqs from NEC}, imply that the spacetime is past-incomplete for null geodesics, assuming the existence of a time when the universe is expanding in all directions.

Before formalizing this, we recall a simple result characterizing completeness of null geodesics~\cite{Sanchez1,Sanchez2}. An elementary fact about Bianchi-I is that each of the vectors
\beq\bal
k_{(1)}^{\mu}&=\left(\frac{c_1}{a_1},\frac{c_1}{a_1^2},0,0\right) \,,\\
k_{(2)}^{\mu}&=\left(\frac{c_2}{a_2},0,\frac{c_2}{a_2^2},0\right) \,,\\
k_{(3)}^{\mu}&=\left(\frac{c_3}{a_3},0,0,\frac{c_3}{a_3^2}\right) \,,\\
\eal\eeq
with $c_i>0$ a set of arbitrary constants, is null (and future-directed) and satisfies the geodesic equation. Consider for definiteness a geodesic with tangent vector $k_{(1)}^{\mu}$ and set $c_1=1$. If the geodesic is past-complete, then it is defined for a semi-infinite range of affine parameter, $\lambda\in(-\infty,\lambda_{\rm f})$. Thus the integral
\beq \label{eq:completeness criterion}
\int_{-\infty}^{t_{\rm f}}a_1(t)\,\d t = \int_{-\infty}^{\lambda_{\rm f}}\d \lambda
\eeq
must diverge (here $t_{\rm f}\equiv t(\lambda_{\rm f})$). The contra-positive statement also follows: convergence of the time integral is equivalent to past-incompleteness. Of course, for this expression to make sense one must assume, as a necessary condition, that the scale factor is positive and defined for all $t\in (-\infty,t_{\rm f})$. In other words, past-completeness requires the absence of curvature singularities, but the converse is not necessarily true; see e.g.~\cite{Witten:2019qhl}.

\textbf{Theorem.} Let the scale factors $a_1,a_2,a_3:(-\infty,t_{\rm f})\to(0,\infty)$ of a Bianchi-I metric be smooth functions and assume
\begin{enumerate}
	\item[\rm(i)] $H_i(t_{\rm f})>0$ for each $i\in\{1,2,3\}$;
	\item[\rm(ii)] the NEC holds.
\end{enumerate}
Then $\int_{-\infty}^{t_{\rm f}} a_i(t)\,\d t < \infty$ for every $i\in\{1,2,3\}$, i.e.\ the spacetime is geodesically past-incomplete.

\textit{Proof.} Define $\tau:=t_{\rm f}-t\in(0,\infty)$ (with primes denoting derivatives with respect to $\tau$) and
\beq
F_i(\tau):=\int_0^\tau H_i(t_{\rm f}-s)\,\d s \,,\qquad S_{jk}:=H_j+H_k \,.
\eeq
It follows that $F_i(0)=0$, $F_i'=H_i$, and $a_i(t_{\rm f}-\tau)=a_i(t_{\rm f})\,e^{-F_i(\tau)}$.

Fix a cyclic triple $(i,j,k)$. The NEC inequality \eqref{eq:H_i ineqs from NEC} may be expressed as $\dot{S}_{jk}\leq \theta H_i-Q$, where $Q:=\sum_i H_i^2$. Using the algebraic identity $\theta H_i-Q=H_iS_{jk}-\frac{1}{2}S_{jk}^2-\frac{1}{2}(H_j-H_k)^2$ we then infer
\beq \label{eq:proof step1}
S_{jk}'\geq S_{jk} \left(\frac{S_{jk}}{2}-H_i\right)+\frac{(H_j-H_k)^2}{2} \,.
\eeq

Define next
\beq
\Psi_i(\tau)
:= S_{jk}(\tau)
\exp\Bigl[F_i(\tau)-\tfrac{1}{2}\bigl(F_j(\tau)+F_k(\tau)\bigr)\Bigr] \,.
\eeq
Its derivative, $\Psi_i'= \Bigl[S_{jk}'+S_{jk}\Bigl(H_i-\tfrac{S_{jk}}{2}\Bigr)\Bigr]
e^{F_i-(F_j+F_k)/2}$, is non-negative as a consequence of \eqref{eq:proof step1}. Thus $\Psi_i(\tau)\geq\Psi_i(0)=: d_i$ for all $\tau\geq0$. Moreover,
\beq
d_i=S_{jk}(t_{\rm f})>0 \,,
\eeq
as per assumption (i). As a by-product, $S_{jk}(\tau)=\Psi_i(\tau)e^{(F_j+F_k)/2-F_i}\geq d_ie^{(F_j+F_k-2F_i)/2}>0$, i.e.\ the pairwise sums are strictly positive.

Rearranging one has
\beq
d_i e^{-F_i(\tau)}\leq S_{jk} e^{-(F_j+F_k)/2}=-2\frac{\d}{\d\tau}\left[e^{-(F_j(\tau)+F_k(\tau))/2}\right] \,.
\eeq
Since both sides of the inequality are positive one may integrate from $\tau=0$ to an arbitrary $\tau=T>0$,
\beq
d_i\int_0^T e^{-F_i(\tau)}\,\d\tau \leq 2\left(1-e^{-(F_j(T)+F_k(T))/2}\right)< 2 \,.
\eeq
Letting $T\to\infty$ one infers the bound
\beq
\int_0^\infty e^{-F_i(\tau)}\,\d\tau < \frac{2}{d_i}<\infty \,,
\eeq
which in turn implies that the integral $\int_{-\infty}^{t_{\rm f}}a_i(t)\,\d t=\int_0^\infty a_i(t_{\rm f}-\tau)\,\d\tau =a_i(t_{\rm f})\int_0^\infty e^{-F_i(\tau)}\,\d\tau$ is convergent.\qed

The proof technique employed here introducing a weighted quantity $\Psi_i$ whose monotonicity follows from the NEC is reminiscent of methods used in the mathematical literature to study the long-time behavior of solutions to geometric flow equations and to establish criteria for geodesic incompleteness~\cite{Galloway:2017mts,Ling:2025slu,Galloway:2025bth}. The construction of such Lyapunov-type functions is a standard tool in the analysis of ordinary differential inequalities arising from energy conditions, and our approach may be of independent interest for other anisotropic spacetimes.

\section{Geodesic completeness and expansion} \label{sec:counterexample}

It is clear that assumption (i) is an essential ingredient in the proof of the theorem. We refer to it as the `present expansion condition', since in the context of our universe we have the basic observational fact that it is expanding in all directions at the present time (and throughout most or perhaps all of its history). It is thus a very natural assumption from a physics perspective.

\begin{figure}
\centering
\includegraphics[width=1\columnwidth]{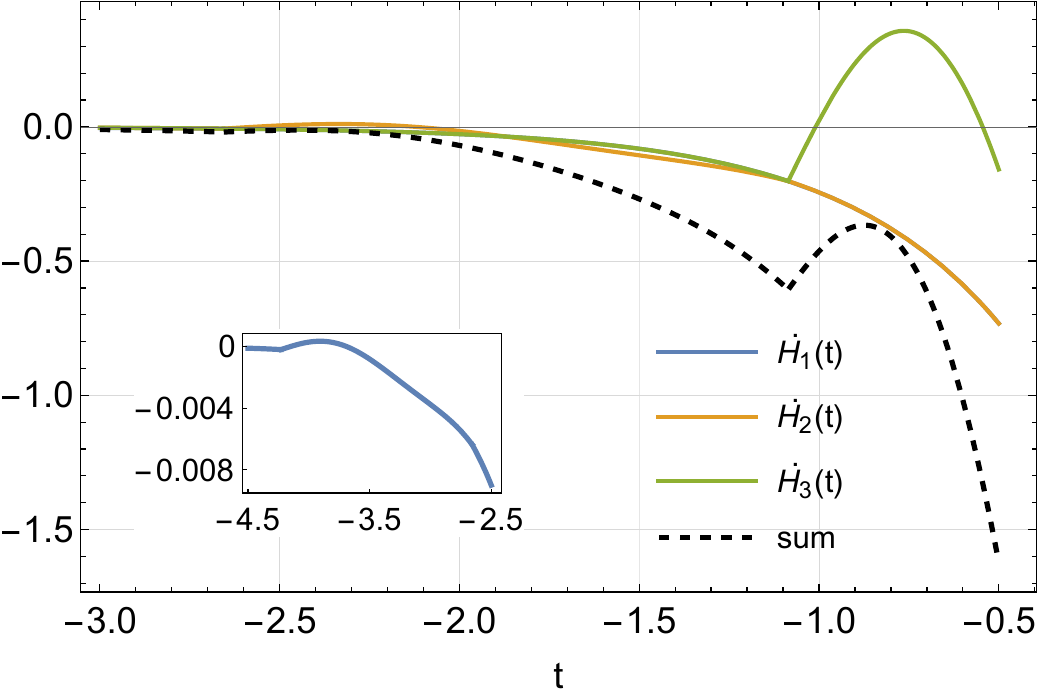}
\caption{Graphs of the time derivatives of the Hubble parameters of the model defined by Eq.\ \eqref{eq:model Hi}. The sum of the three is also included, which is seen to be everywhere negative, consistent with the NEC. The inset shows the graph of $\dot{H}_1$, which is hidden by the other curves in the full plot.}
\label{fig:nec1}
\end{figure}

\begin{figure}
\centering
\includegraphics[width=1\columnwidth]{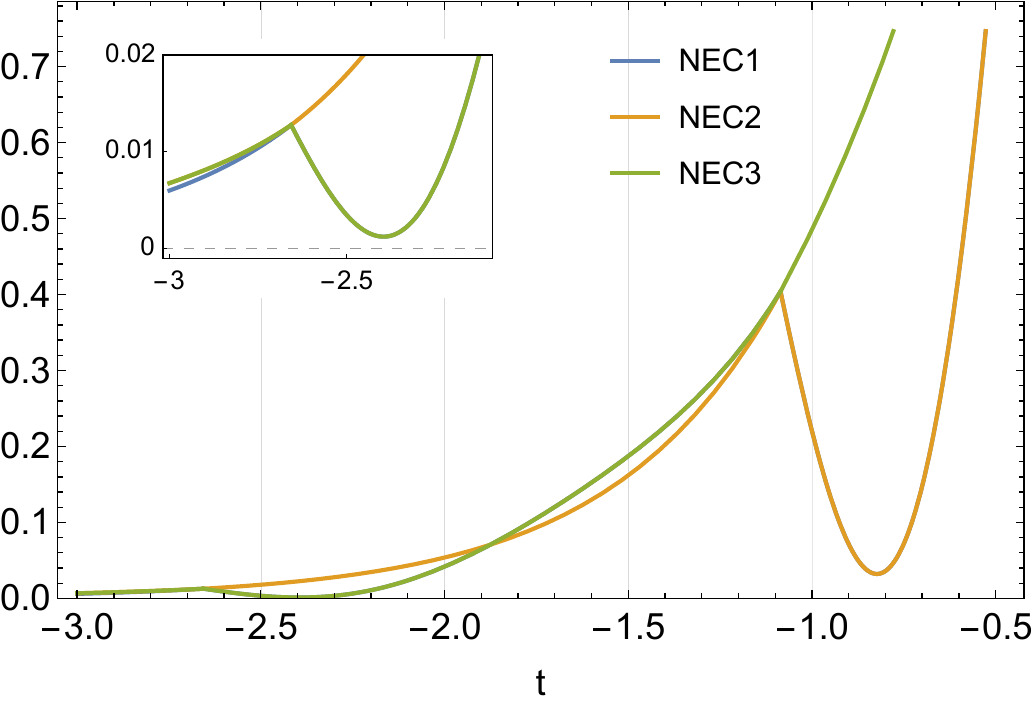}
\caption{Graphs of the left-hand sides of the three NEC inequalities given in \eqref{eq:H_i ineqs from NEC} for the model defined by Eq.\ \eqref{eq:model Hi}. The three curves are seen to be everywhere positive. The inset is a zoomed-in part of the plot. The graph of the first inequality, barely visible in the inset, is mostly hidden by the other curves.}
\label{fig:nec2}
\end{figure}

It is nevertheless interesting to ask whether the present expansion condition is indeed necessary for the conclusion of the theorem to follow. The answer is yes, and in fact we have constructed an explicit example of a Bianchi-I spacetime which is geodesically past-complete and consistent with the NEC, but does not satisfy assumption (i), i.e.\ it is always contracting along some direction, but each direction exhibits a phase of expansion. It is defined by the following set of Hubble parameters:\footnote{The functions $H_i$ in this example are $C^1$ but not smooth. There is however no difficulty in constructing a smooth version of this model by use of a mollifier.}
\beq \label{eq:model Hi}
H_i(t)=\begin{cases}
	-e^{At}+B_i\left[1+\cos\left(\omega(t-t_i)\right)\right] \\
	\qquad\qquad \mbox{if $t\in\left(t_i-\pi/\omega,t_i+\pi/\omega\right)$}\,, \\
	-e^{At} \\
	\qquad\qquad \mbox{otherwise} \,,
\end{cases}
\eeq
and the choice of parameters (in arbitrary units): $A=2.2$, $B_i=\{0.000195,0.0062,0.2\}$, $\omega=4$, $t_i=\{-0.3-4\pi/\omega,-0.3-2\pi/\omega,-0.3\}$. Fig.\ \ref{fig:nec1} displays the graphs of $\dot{H}_i$, showing that, while each attains positive values, their sum remains negative at all times, consistent with the NEC. This is further illustrated in Fig.\ \ref{fig:nec2} which shows that the three NEC inequalities, Eq.\ \eqref{eq:H_i ineqs from NEC}, are indeed satisfied at all times. The reader may also check that all three $H_i$ attain positive values in different time intervals.

It is also interesting to inquire about the relationship between our result and the BGV theorem mentioned in the Introduction. We recall a result that follows from the application of the BGV theorem to a geodesic congruence with tangent vector along one of the Cartesian axes~\cite{Garcia-Saenz:2024ogr}:\footnote{This result was established in Ref.~\cite{Garcia-Saenz:2024ogr} in the context of FLRW spacetimes. The proof however also applies in the case of Bianchi-I upon considering the particular case of a geodesic congruence directed along one of the Cartesian axes.} suppose the spacetime is past-complete and that there exists a time $t_i$ such that $H_i(t_i)>0$; then there exists an interval $\mathcal{I}_i\subset (-\infty,t_i)$ such that $\dot{H}_i(t)>0$ for all $t\in \mathcal{I}_i$. Notice that, in contrast to (i), here the times $t_i$ are not assumed equal. This result immediately implies a violation of the NEC in the case of FLRW, but not in the case of Bianchi-I. The reason is that the three intervals $\mathcal{I}_i$ need not overlap, so it not necessary for the sum $\dot \theta=\sum_i\dot{H}_i$ to attain positive values, even if the individual $\dot{H}_i$ must do so in some time interval. The above counter-example is indeed consistent with this observation.


\section{Discussion} \label{sec:discussion}

We have demonstrated a theorem establishing that Bianchi-I cosmological spacetimes must be geodesically past-incomplete within GR if matter satisfies the NEC and provided the universe is expanding in all directions at some time. While specific to Bianchi-I metrics, and hence not at the level of generality of other `singularity' theorems, our result has the virtue of manifestly relying on plausible physical conditions, in particular without any assumptions on the global topology of the spacetime. Also interesting, we believe, is the observation that the NEC by itself is not sufficient to conclude geodesic incompleteness. To the best of our knowledge, the example of a past-complete Bianchi-I metric that we constructed, characterized by exhibiting expansion in each direction, is unprecedented. Although admittedly contrived, it could serve as the starting point for a general classification of spacetimes of this kind.

Let us comment on some potential extensions of our analysis. First, our theorem strictly speaking concerns null geodesic completeness. We see no significant obstacle in running the argument for time-like geodesics, so we expect the same conclusion to follow in this case too. Second, we already mentioned that the assumption on the number of spatial dimensions is inessential, so again we expect the analysis to apply in a rather verbatim fashion to Bianchi-I metrics of arbitrary dimension. We also remark that our assumption that the Einstein equation holds was used only to relate the NEC with the null convergence condition (NCC), which is a purely geometric statement. Our theorem therefore applies in any metric theory of gravity if one replaces the NEC with the NCC.

As for the limitations, it must be admitted that our study relies strongly on the form of the Bianchi-I metric. Specifically, several steps in our analysis leverage results that have been established for flat FLRW metrics. It is therefore unclear if our techniques may apply to other anisotropic spacetimes, not to mention inhomogeneous ones. We hope however that our results will serve at least as a useful conceptual precedent, underscoring the assumptions that one might need to establish more general theorems.

As a final remark, it is worth recalling that the NEC is not truly fundamental at the quantum level, although it is presumably an essential requirement for classical fields. The condition that may actually be demonstrated within quantum field theory is the averaged null energy condition (ANEC), which states that the integral of $T_{\mu\nu}l^{\mu}l^{\nu}$ along a null geodesic with tangent vector $l^{\mu}$ must be non-negative, while allowing for the integrand to be transiently negative. Recently the ANEC has been applied to FLRW cosmologies to establish that only spacetimes with positive spatial curvature can be geodesically complete~\cite{Burwig:2025hrr,Burwig:2026fsy}. It would be interesting to reassess our study of anisotropic cosmologies in light of this result.

\vskip 5pt

\textit{Acknowledgments.---}We are grateful to Yunke Zhao for collaboration in the early stages of this project. SGS acknowledges support from a Provincial Grant (Grant No.\ 2023QN10X389); he would also like to thank the Henan Academy of Sciences for generous hospitality. AS acknowledges that this research is supported by the Institute of Mathematics, funded through the High-level Talent Research Start-up Project Funding of the Henan Academy of Sciences (Project No.: 251819085).

\appendix

\section{General form of the null energy condition} \label{app:general NEC}

Eq.\ \eqref{eq:H_i ineqs from NEC} in the main text follows from applying the NEC along each of the Cartesian directions. More generally, we may consider a general null vector $l^{\mu}$. In an orthonormal frame with vielbein $e_{\hat{\mu}}{}^{\mu}$ we define
\beq
l^{\hat{\mu}}=\left(1,n_1,n_2,n_3\right) \,, \qquad \sum_i n_i^2=1 \,.
\eeq
Then,
\beq
R_{\mu\nu}l^{\mu}l^{\nu}=R_{\mu\nu}e_{\hat{\mu}}{}^{\mu}e_{\hat{\nu}}{}^{\nu}l^{\hat{\mu}}l^{\hat{\nu}}=R_{00}+\sum_{i,j}R_{ij}\frac{n_in_j}{a_ia_j} \,,
\eeq
where $R_{\mu\nu}$ is the Ricci tensor and we used that $R_{0i}=0$ for the Bianchi-I metric. Furthermore,
\beq\bal
R_{00}&=-\sum_i \frac{\ddot{a}_i}{a_i}=-\sum_i \dot{H}_i+H_i^2 \,,\\ R_{ij}&=a_i^2\left[\dot{H}_i+\theta H_i\right]\delta_{ij} \qquad \mbox{(no sum)}\,.
\eal\eeq
The NEC $R_{\mu\nu}l^{\mu}l^{\nu}\geq0$ is then equivalent to
\beq \label{eq:H_i ineqs general direction}
\sum_i \left[-(1-n_i^2)\dot{H}_i-H_i^2+\theta H_in_i^2\right]\geq0 \,.
\eeq
Choosing e.g.\ $\vec{n}=(1,0,0)$ or along any of the Cartesian axes reproduces the inequalities in \eqref{eq:H_i ineqs from NEC}.

We remark however that \eqref{eq:H_i ineqs general direction} does not contain more information than \eqref{eq:H_i ineqs from NEC}. The reason is that \eqref{eq:H_i ineqs general direction} is linear in the quantities $n_i^2$, and the constraint $\sum_i n_i^2=1$ defines a convex polytope in $\mathbb{R}^3$. It follows that the minimum of the left-hand side, at each time $t$, must be attained at one of the vertices, which is nothing but the result of applying the NEC along one of the Cartesian axes.\footnote{Notice however that it is not sufficient to consider a single vertex because the minimizing direction may change with $t$.}

\section{Application of the Raychaudhuri equation} \label{app:raych}

In the main text we derived the bound $\dot{H}\leq0$ from the Einstein equations and the NEC. It is also instructive to observe that this also follows from the Raychaudhuri equation. Define
\beq
S_i:=H_{i+1}+H_{i+2} \,,
\eeq
understood again in a cyclic sense, i.e.\ with $H_{4}\equiv H_1$ and $H_{5}\equiv H_2$. Notice that \eqref{eq:H_i ineqs from NEC} implies
\beq \label{eq:raych S_i}
-\dot{S}_i+S_iH_i-\frac{1}{2}S_i^2\geq0 \quad \forall\,i \,,
\eeq
since $-\frac{1}{2}S_i^2\geq -(H_{i+1}^2+H_{i+2}^2)$.

Consider a null geodesic congruence along the $x^1$ axis. Recall that $k^{\mu}=\left(\frac{c}{a_1},\frac{c}{a_1^2},0,0\right)$, where $c>0$ is an integration constant. The expansion is then $\theta_1:=\nabla_{\mu}k^{\mu}=\frac{c}{a_1}S_1$. The null Raychaudhuri equation together with the NEC implies (see e.g.~\cite{Hawking:1973uf})
\beq
\frac{\d\theta_1}{\d\lambda}\leq -\frac{1}{2}\theta_1^2 \,,
\eeq
for any null congruence. Dividing by $\d t/\d\lambda$ to get a time derivative then yields \eqref{eq:raych S_i} for $i=1$. Repeating this argument for the other axes shows that \eqref{eq:raych S_i} follows from the Raychaudhuri equation. We remark that this is a weaker condition than the result obtained directly from the Einstein equation.

Summing over the inequalities \eqref{eq:raych S_i} we get
\beq
-2\dot{\theta}-\frac{1}{2}\sum_{i=1}^3(H_i-H_{i+1})^2 \geq0 \,,
\eeq
and therefore $\dot{\theta}\leq0$, with equality only possible in the isotropic case.

\bibliographystyle{apsrev4-1}
\bibliography{BianchiCompleteness}
	
\end{document}